\RequirePackage{fix-cm}

\documentclass[twocolumn]{svjour3}  
\smartqed 

\usepackage{graphicx}% Include figure files
\usepackage[misc]{ifsym}

\graphicspath{{./figures/}}

\begin{document}

\title{External and mutual synchronization of chimeras in \\a two layer network of nonlinear oscillators}%

\titlerunning{External and mutual synchronization of chimeras}   

\author{Andrei V. Bukh \and Galina I. Strelkova \and \\Vadim S. Anishchenko}

\institute{Andrei V. Bukh \and Galina I. Strelkova \and \\ Vadim S. Anishchenko (\Letter) \at
              Department of Physics, Saratov State University, \\ 
              Astakhanskaya str. 83, 410012 Saratov, Russia \\
              \email{buh.andrey@yandex.ru}           %  \\
           \and
           Galina I. Strelkova \at
           \email{strelkovagi@info.sgu.ru}
          \and
           Vadim S. Anishchenko   \at       
           \email{wadim@info.sgu.ru}
}
\date{Received: date / Accepted: date}

\maketitle   

\begin{abstract}

We study numerically synchronization phenomena of spatiotemporal structures, including chimera states, in a two layer network of nonlocally coupled nonlinear chaotic discrete-time systems. Each of the interacting ensembles represents a one layer ring network of nonlocally coupled logistic maps in the chaotic regime. The coupled networks differ in their control parameters that enables one to observe distinct spatiotemporal dynamics in the networks  when there is no coupling between them. We explore in detail external and mutual synchronization of chimera structures. The identity of synchronous structures and the estimation of synchronization regions are quantified by calculating the cross-correlation coefficient between relevant oscillators of the networks. We show that for non-identical networks, unidirectional and symmetric couplings lead to external and mutual synchronization between the interacting ensembles, respectively. This is confirmed by identical synchronous structures and by the existence of finite regions of synchronization within which the cross-correlation coefficient is equal to 1.  We also show that these findings are qualitatively equivalent to the results of the classical synchronization theory of periodic self-sustained oscillations.

%\pacs{05.45.-a, 02.60.-x}% PACS, the Physics and Astronomy
%Use showkeys class option if keyword

\end{abstract}

\section{Introduction}
\label{part_intro}

Exploring the collective dynamics of complex systems of different nature, synchronization of 
ensembles of interacting oscillators and dissipative structure formation are (occupy) in the focus ( in the center) of attention  of many researchers in nonlinear dynamics during already the past few ten years. 
A lot of monographs \cite{AFR95,NEK02,OSI07,PIK01} and papers 
\cite{Nek-Mak95,Nek-Vel97,Nek-Vel99,Belykh00,Belykh01,Vad05,Pecora14} were devoted to this research topic.  It has been established that nonlinear ensembles and networks can typically demonstrate  structure formation, such as synchronization clusters, spatial intermittency, spatially stationary (motionless)  regular and chaotic structures, etc. The majority of works deal, as a rule, with ensembles of identical oscillatory elements with local or global coupling between them. Recently, a novel type of spatio-temporal structures, called a chimera state \cite{KUR02},  has been discovered in ensembles of nonlocally coupled identical phase oscillators. Nonlocal coupling means that each ensemble oscillator  is connected with a finite number of neighbors from the left and right. The chimera state has been initially revealed in a network of nonlocally coupled identical phase oscillators  \cite{KUR02} and then explored in detail in the paper  \cite{ABR04,Panaggio-Abrams2015} where the term "chimera state" has been proposed. The chimera  denotes such a dynamical state when the network splits into coexisting domains with synchronized (coherent) and desynchronized (incoherent) behavior of the network elements and these clusters are clearly spatially localized.    

We note that chimera-like structures were observed long before their definition in the paper \cite{ABR04}.
The attention of researchers in those years was aimed at the analysis of transitions from spatiotemporal chaos to complete synchronization of ensemble elements \cite{Waller84,Kaneko89}. Chimera-like patterns were found to be transient to the synchronization regime and were called as synchronous and desynchronous clusters (see, for example, \cite{Astakhov95,Springer07}). Recently, the discovery of chimera states attracted much attention, aroused a great interest of many researchers and led to the growth of numerical and theoretical \cite{Abrams08,Laing10,Laing11,Martens10,Motter10,Wolfrum11,OME2011,OME2012,Maistrenko14,Yeldesbay14,EPL2015,Chaos2016,Krischer2016,PRL2016,Schoell2016,Chaos-int2017,CNSNS2017} and then experimental 
\cite{Hagerstrom-NaturePhys-2012,Tinsley-NaturePhys-2012,Larger-PRL-2013,Martens13,Kapitaniak14,Larger15} studies. The analysis of various spatiotemporal patterns (including chimeras) in complex ensembles has not only a fundamental scientific but also important practical significance. This is particularly important in exploring  arrays of Josephson junctions \cite{Watanabe95}, large arrays of coupled lasers \cite{Li94}, neural networks \cite{Hizanidis16}, brain dynamics \cite{RAT00}, power grids \cite{MOT13,Nishikawa15}, etc. 

Real-world networks, however, are typically not isolated and always functionally and conditionally connected to other networks. It is therefore important to explore the dynamics of coupled or multilayer networks \cite{BOC14,Majhi16,Maksimenko16,Jalan16,Perc17}.
 In this case synchronization of various spatiotemporal patterns in such systems becomes one of the main and intriguing research topics.  This problem was studied before for coupled dynamical systems
 \cite{Nek-Vel97,Pecora90,Rulkov95,Ros96,Kocarev96} but basically for coupled identical oscillators with local or global coupling. Moreover, synchronous structures and synchronization regions were not quantitatively estimated  and defined in the parameter space. 
Recently, a number of works have appeared which are devoted to synchronization between coupled complex networks \cite{Li07,Tang08,Wu09,Wu12,Andrj-Chaos17,ChaosFT17}. 
However, there are only few papers where it has been shown that chimera states can be synchronized across networks \cite{Andrj-Chaos17,ChaosFT17}. 
%in coupled networks of nonlocally coupled phase oscillators (identical and generalized synchronization) \cite{Andrj-Chaos17} and of non-identical chaotic maps (mutual synchronization) \cite{ChaosFT17}. 

 In our work  we aim to study effects of external and mutual synchronization of chimera states and spatiotemporal structures in a two layer network made of coupled ensembles of nonlocally coupled chaotic discrete-time systems. The individual elements  are described by the well-known logistic map  with a control parameter detuning (mismatch) in each sub-network. The logistic map is the simplest and famous example of a wide class of nonlinear chaotic systems with the Feigenbaum scenario of  chaos onset. Moreover, it has been shown that an ensemble of nonlocally coupled logistic maps in the chaotic regime can realize chimera states \cite{OME2011,OME2012}. The control parameter detuning, which we introduce in our study,  enables us to implement various spatiotemporal structures, including chimeras, when the one layer networks are uncoupled.  We quantify the identity degree of synchronous patterns by using the cross-correlation coefficient between relevant elements of the coupled ensembles. This quantity is also applied to calculate and estimate  synchronization regions in different parameters' plane.

\section{External synchronization of chimera structures}

\begin{figure}[htbp]
\begin{center}
\includegraphics[width=.7\columnwidth]{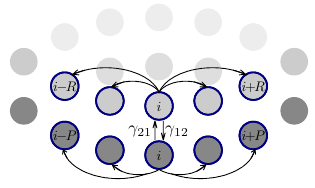}
\end{center}
\caption{Schematical representation of a two layer multiplexing network. Each ring sub-network consists of nonlocally coupled oscillators. The $i$th oscillators ($i=1,2,\ldots, N$) of the interacting rings are coupled with coupling coefficients $\gamma_{21}$ and $\gamma_{12}$}
\label{two_coupled_rings}                                                                                                   
\end{figure}

We consider the dynamics of the two layer multiplexing network consisting of nonlocally coupled nonlinear chaotic oscillators, which is schematically shown in Fig.~\ref{two_coupled_rings}.

%%%%%%%%Figure 1%%%%%%%%%%%%%%%%%%%%%%%%%%%%%%%%%%%%%%%%%
%%%%%%%%%%%%%%%%%%%%%%%%%%%%%%%%%%%%%%%%%%%%%%%%%%

The model under study is described by the following system of equations:
\begin{eqnarray}\label{main_eq}
%\begin{aligned}
   x_i^{t+1} &=& f_i^t + \frac{\sigma_1}{2P} \sum_{j=i-P}^{i+P} \left[ f_j^t - f_i^t\right] + \gamma_{21} F_i^t, \nonumber \\ y_i^{t+1} &=& g_i^t + \frac{\sigma_2}{2R} \sum_{j=i-R}^{i+R} \left[ g_j^t - g_i^t\right] + \gamma_{12} G_i^t,
%\end{aligned}
\end{eqnarray}
where $x_i^t$ and $y_i^t$ are real dynamical variables of coupled ensembles, $i$ is the number of the ensemble element, $i=1,2,\ldots, N$, $N$ is the total number of elements in each sub-network, $t$ denotes the discrete time. In our study the individual elements in both ensembles are defined by the logistic map $f_i^t=\alpha_1 x_i^t(1-x_i^t)$ and $g_i^t=\alpha_2 y_i^t(1-y_i^t)$ with different control parameters  $\alpha_{1}$ and $\alpha_{2}$.  $\sigma_1$ and $\sigma_2$ specify the nonlocal coupling strengths in the rings (intra-couplings), $P$ and $R$ denote the number of neighbors of the $i$th element from each side in the first and second sub-systems, respectively. $F_i^t=(g_i^t-f_i^t)$ and $G_i^t=(f_i^t-g_i^t)$ are the coupling functions between the sub-networks (inter-coupling) and $\gamma_{12}$ and $\gamma_{21}$ are the inter-coupling strengths. 

In order to analyze synchronization effects we introduce a control parameter detuning in both rings by setting 
 $\alpha_1=3.7$ and $\alpha_2=3.85$ which correspond to the chaotic regime in the individual elements of both ensembles. We also choose different intra-couplings between the elements in each ring as  $\sigma_1=0.23$ and $\sigma_2=0.15$. We fix the equal number of  neighbors $P=R=320$, at which the uncoupled networks show chimera states. Each sub-network consists of $N=1000$ elements.  
Equations~(\ref{main_eq}) are solved numerically  for periodic boundary conditions (we consider the ring network scheme) and for the initial conditions  $(x_i^0,y_i^0)$ randomly distributed across the network nodes  $i=1,2,\dots,N$ within the interval $[0;1]$. The iteration time is $10^5$ and the transient of $5\times 10^5$ iterations is discarded.
 
 We start with considering external synchronization in the model (\ref{main_eq}). In this case 
 the inter-coupling is unidirectional, i.e., $\gamma_{21}=0$ and  $\gamma_{12}=\gamma>0$. This means that nodes in the first  (driver) network are coupled unidirectionally to nodes in the second (response) network. A similar scheme of the drive-response network pairs was considered in \cite{Li07,Tang08,Andrj-Chaos17}.

\begin{figure}[htbp]
    \centering
    \includegraphics[width=0.37\textwidth]{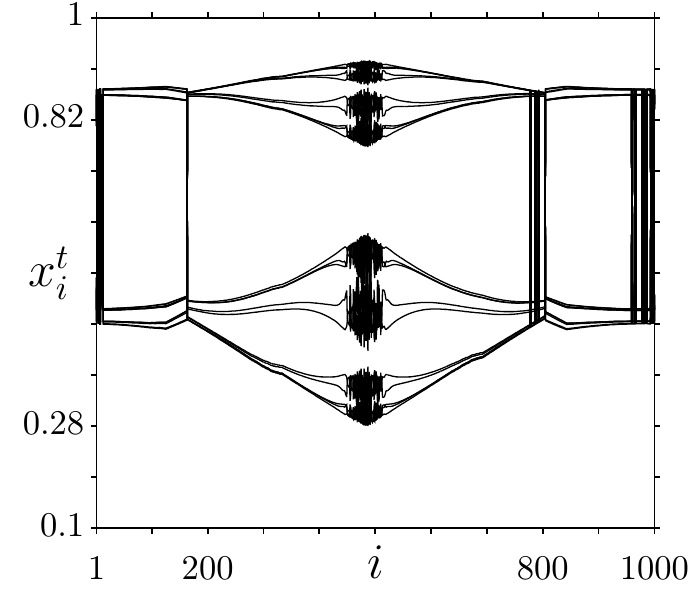}
    \\[-10.3em]~\hspace{21.5em}(a)
    \\[8.5em] \includegraphics[width=0.37\textwidth]{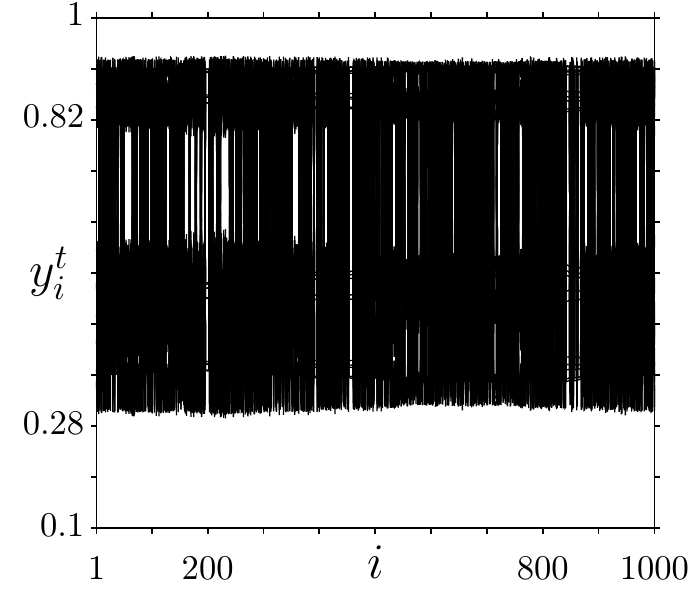}
    \\[-10.3em]~\hspace{21.5em}(b)
    \\[8.5em] \caption{Spatiotemporal profiles of amplitudes $x_i^t$ (a) and $y_i^t$ (b) in the network (\ref{main_eq}) without coupling ($\gamma=0$) and for  $\alpha_1=3.7$, $\alpha_2=3.85$, $\sigma_1=0.23$, and $\sigma_2=0.15$}
    \label{pic2_without_coupling}
\end{figure}

As the considered sub-networks differ in the control parameter of their individual nodes ($\alpha_1=3.7,\alpha_2=3.85$),  they can show individually various spatiotemporal patterns, including chimera states,  when there is no coupling between them. Typical structures are exemplified in Fig.~\ref{pic2_without_coupling} by  spatiotemporal profiles for the amplitudes $x_i^t$ and $y_i^t$ of each ring. 
A spatiotemporal profile \cite{CNSNS2017} represents the 100 last iterations for amplitudes (coordinates) of each network elements and thus visualizes the temporal dynamics of the whole network.

%%%%%%%%Figure 2%%%%%%%%%%%%%%%%%%%%%%%%%%%%%%%%%%%%%%%%%
%%%%%%%%%%%%%%%%%%%%%%%%%%%%%%%%%%%%%%%%%%%%%%%%%%

\begin{figure}[htbp]
    \centering
    \includegraphics[width=0.37\textwidth]{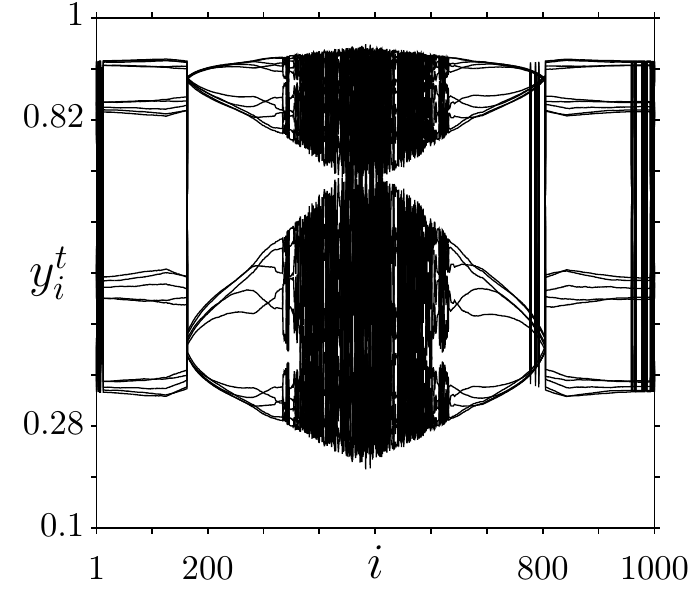}
    \\[-10.3em]~\hspace{21.5em}(a)
    \\[8.5em] \includegraphics[width=0.37\textwidth]{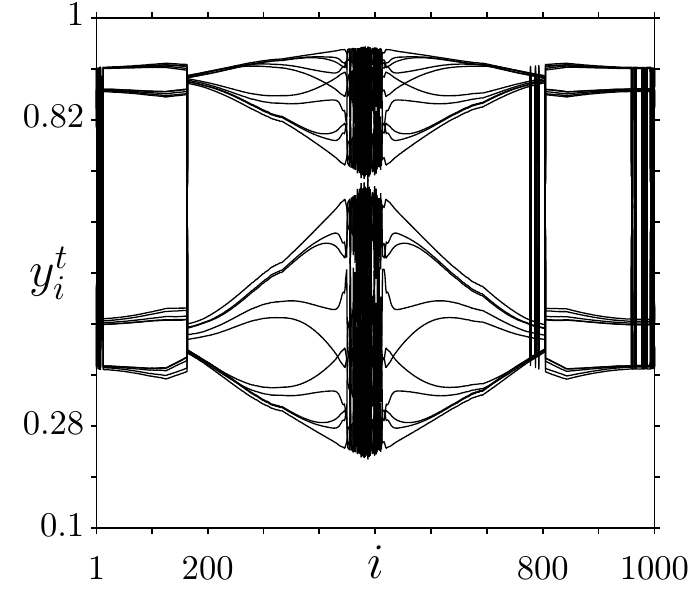}
    \\[-10.3em]~\hspace{21.5em}(b)
    \\[8.5em] \caption{Spatiotemporal profiles of dynamical states $y_i^t$ of the response network in (\ref{main_eq}) when affected by the driver with the inter-coupling strength  $\gamma=0.15$ (a) and $\gamma=0.3$ (b). Other parameters are as in Fig.~\ref{pic2_without_coupling} }
    \label{pic3_non_synced}
\end{figure}

As can be seen, the driver network $x_i^t$ shows amplitude and phase chimeras \cite{OME2011,CNSNS2017} (Fig.~\ref{pic2_without_coupling},a), and the second ring $y_i^t$ demonstrates spatiotemporal chaos (Fig.~\ref{pic2_without_coupling},b). When the unidirectional coupling with strength $\gamma>0$  is introduced, the influence of the driver network induces the appearance of amplitude and phase chimeras in the response network. However, if the inter-coupling is sufficiently weak, e.g., $\gamma=0.15$ and $\gamma=0.3$, the spatiotemporal profile of $y_i^t$ does not coincide yet completely with that one for $x_i^t$ at $\gamma=0$ (Fig~\ref{pic2_without_coupling},a). The corresponding results for the dynamics of the response network $y_i^t$ are shown in Fig.~\ref{pic3_non_synced}.

%%%%%%%%% Figure  3%%%%%%%%%%%%%%%%%%%%%%%%%%%%%%%%%%%%%%%%%
%%%%%%%%% 3%%%%%%%%%%%%%%%%%%%%%%%%%%%%%%%%%%%%%%%%%

\begin{figure}[htbp]
    \centering
    \includegraphics[width=0.37\textwidth]{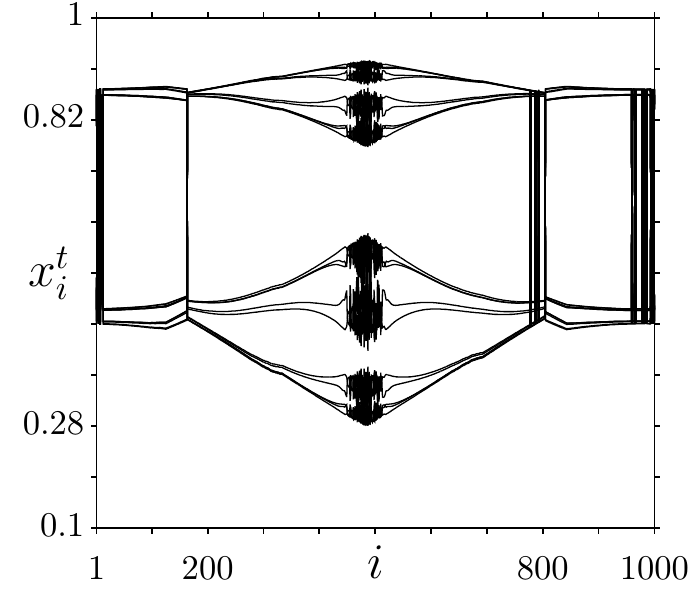}
    \\[-10.3em]~\hspace{21.5em}(a)
    \\[8.5em] \includegraphics[width=0.37\textwidth]{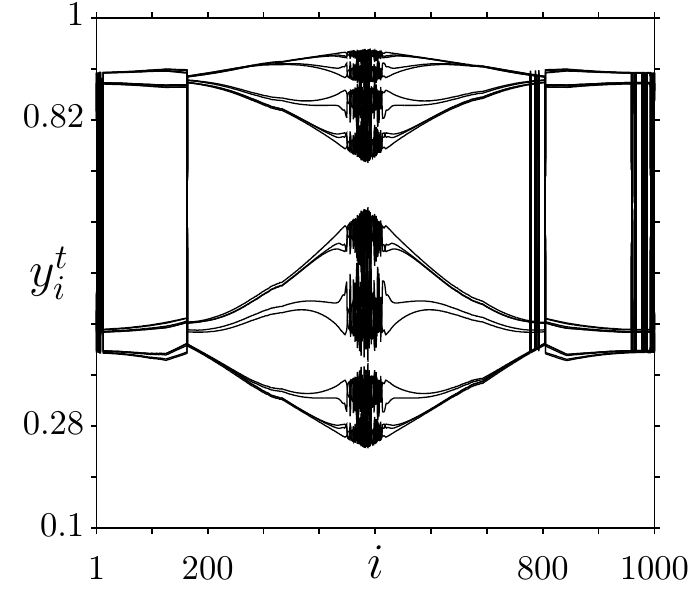}
    \\[-10.3em]~\hspace{21.5em}(b)
    \\[8.5em] \caption{Spatiotemporal profiles of the $x_i^t$ (a) and $y_i^t$ (b) coordinates of the network  (\ref{main_eq})  in the synchronization mode at $\gamma=0.45$. Other parameters are as in Fig.~\ref{pic2_without_coupling}}
    \label{pic4_synced}
\end{figure}

If the inter-coupling strength increases further,  $\gamma\geq 0.4$, the structures in both rings become rather identical (Fig.~\ref{pic4_synced}) and  there are only minor differences in their amplitudes (coordinates' values). 

%%%%%%%%Figure 4 %%%%%%%%%%%%%%%%%%%%%%%%%%%%%%%%%%%%
%%%%%%%%%%%%%%%%%%%%%%%%%%%%%%%%%%%%%%%%%%%%%%%%%

\begin{figure}[htbp]
    \centering
    \includegraphics[width=0.37\textwidth]{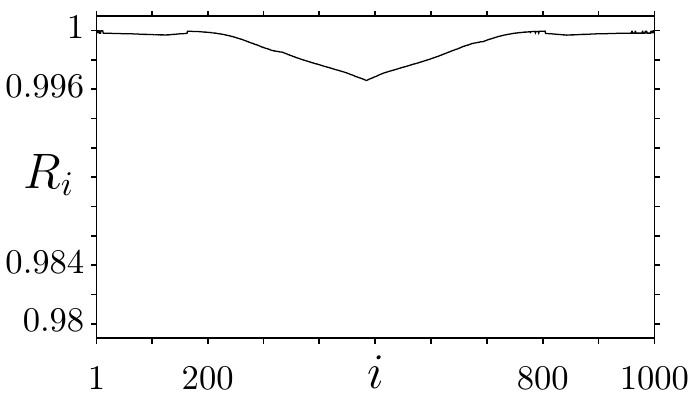}
    \caption{Cross-correlation coefficient $R_i$~(\ref{correlations_eq}) calculated for the structures in  Fig.~\ref{pic4_synced} at $\gamma=0.45$. Other parameters are as in Fig.~\ref{pic2_without_coupling}}
    \label{pic5_correlation}
\end{figure}

In order to conclude that the external synchronization takes place indeed, two important facts must be justified. We need to quantify the identity of synchronized structures (Fig.~\ref{pic4_synced},a,b) and to show that there is a finite  region of synchronization in the system (\ref{main_eq}) parameter plane. This can be done by calculating the cross-correlation coefficient $R_i$ between relevant elements of the coupled networks as follows:
\begin{eqnarray}\label{correlations_eq}
&&R_i = \frac{\langle\tilde x_i(t)\tilde y_i(t)\rangle}{\sqrt{\langle\tilde x_i^2(t)\rangle \langle\tilde y_i^2(t)\rangle},}\quad \\ [.2cm]
%\begin{eqnarray}
&& \tilde{x}_i(t) =  x_i(t)-\langle x_i(t)\rangle, \nonumber \\
&& \tilde{y}_i(t) = y_i(t)-\langle y_i(t)\rangle. \nonumber 
%\end{eqnarray}
\end{eqnarray}
The angle brackets $\langle\cdot\rangle$ in~(\ref{correlations_eq}) mean the time averaging. If spatiotemporal  structures are identical, the cross-corre\-la\-tion coefficient $R_i$ must be equal to 1. The numerical results for $R_i$ are plotted in Fig.~\ref{pic5_correlation} for the regimes presented in Fig.~\ref{pic4_synced},a and b. It is clearly seen that  $0.98<R_i\leq 1.0$ for all $i=1,2,\dots,N$. This implies that the observed structures are identical or synchronous. 

%%%%%%%%Figure 5%%%%%%%%%%%%%%%%%%%%%%%%%%%%%%%%%%%%%%%%%
%%%%%%%%% 5%%%%%%%%%%%%%%%%%%%%%%%%%%%%%%%%%%%%%%%%

\begin{figure}[htbp]
    \centering
    \includegraphics[width=0.37\textwidth]{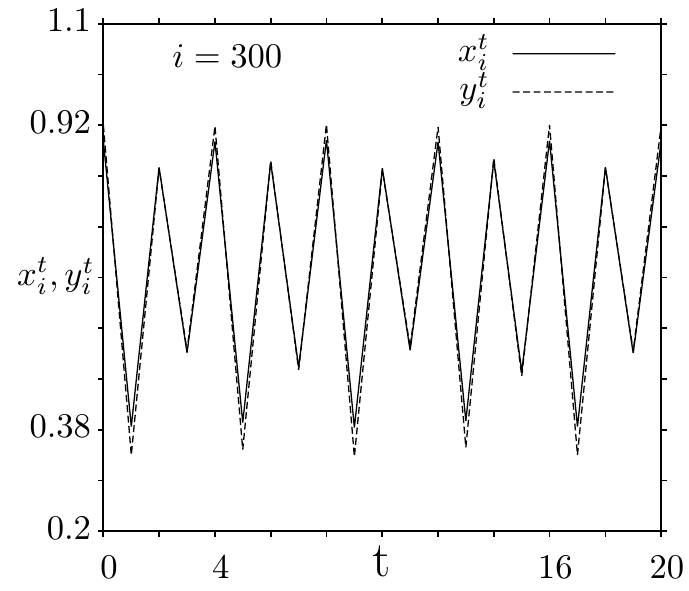}
    \caption{Time series   $x_i^t(t)$ and $y_i^t(t)$  for the $300$th oscillators, calculated in the synchronization mode in the network ~(\ref{main_eq})  at $\gamma=0.45$. Other parameters are as in Fig.~\ref{pic2_without_coupling}}
    \label{pic6_timeseries}
\end{figure}

The synchronization effect  
(Figs.~\ref{pic4_synced} and \ref{pic5_correlation}) can also be well illustrated by computing  time series $x_i^t(t)$ and $y_i^t(t)$ for relevant (symmetric) elements of the coupled sub-networks. The corresponding numerical results are presented in Fig.~\ref{pic6_timeseries} for the $i=300$th oscillators of the model (\ref{main_eq}). As follows from the plots, the oscillations  are rather synchronous and are characterized by only a minor difference in the amplitude values, that is quite acceptable for synchronous regimes.

%%%%%%%%Figure 6%%%%%%%%%%%%%%%%%%%%%%%%%%%%%%%%%%%%%%%%%
%%%%%%%%%%%%%%%%%%%%%%%%%%%%%%%%%%%%%%%%%%%%%%%%%

\section{Regions of external synchronization of spatiotemporal structures}

We now intend to show that the effect of external synchronization can be observed within finite ranges of the system parameter variation. In order to construct  synchronization regions of spatiotemporal structures in the network (\ref{main_eq}) we consider two different cases. Firstly, we leave the control parameter in the driver network  $\alpha_1=3.7$ fixed and vary the control parameter $\alpha_2$ in the response network. When the considered networks are uncoupled, varying  $\alpha_2$ can result in the appearance of various spatiotemporal patterns in the second ring, including chimera states and periodic structures.

\begin{figure}[htbp]
    \centering
    \includegraphics[width=0.37\textwidth]{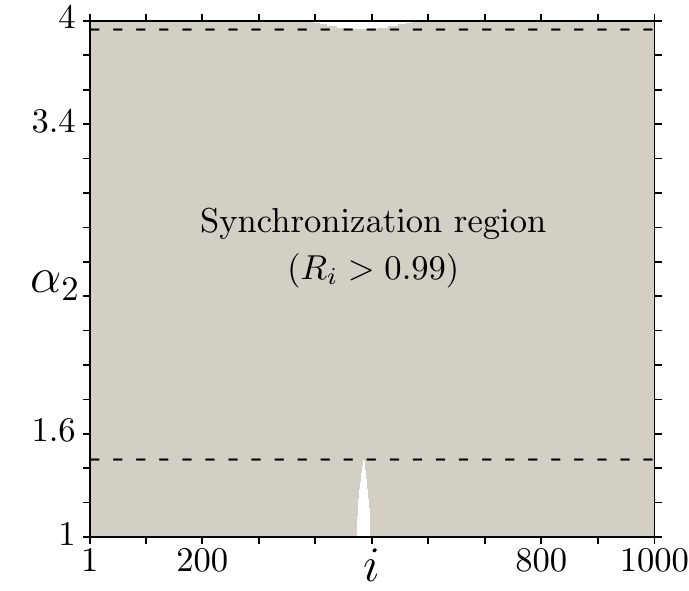}
    \caption{External synchronization region when varying parameter  $\alpha_2$ and for $\gamma=0.4$, $\alpha_1=3.7$, $\sigma_1=0.23$, and $\sigma_2=0.15$ }
    \label{pic7_syncarea}
\end{figure}

Let us analyze what happen with the above mentioned patterns when the driver network $x_i^t$ influences unidirectionally  the response network $y_i^t$ with $\gamma>0$ and when $\alpha_2$ is varied. Our calculations show that at certain values of $\gamma$, the driver network can induce a structure in the response network, which completely coincides with that one in the driver ensemble (Fig.~\ref{pic4_synced}). The resulting synchronous regime is stable and is stored in a finite region of synchronization when the control parameter $\alpha_2$ is varied in the interval $[1.4; 3.9]$.  The corresponding numerical results for the synchronization region is shown in Figure~\ref{pic7_syncarea} for the considered case.

%%%%%%%%Figure 7%%%%%%%%%%%%%%%%%%%%%%%%%%%%%%%%%%%%%%%%%
%%%%%%%%%%%%%%%%%%%%%%%%%%%%%%%%%%%%%%%%%%%%%%%%%

As follows from Fig.~\ref{pic7_syncarea}, at $\gamma=0.4$ there is a finite range (grey color) of parameter $\alpha_2$ 
values  ($1.45\le\alpha_2\le 3.9$), within which the cross-correlation  $R_i>0.99$ for all the elements of the network (\ref{main_eq}). Inside this domain the response network shows the spatiotemporal pattern which is completely identical or synchronous to the structure in the driving network. The state $y_i^t$ of the response sub-system is fully determined  by the driver  dynamics $x_i^t$, i.e, the  external synchronization is clearly observed.  The white color region in Fig.~\ref{pic7_syncarea} corresponds to the lack of synchronization ($R_i<1$) of a part of the network elements. 

\begin{figure}[htbp]
    \centering
    \includegraphics[width=0.37\textwidth]{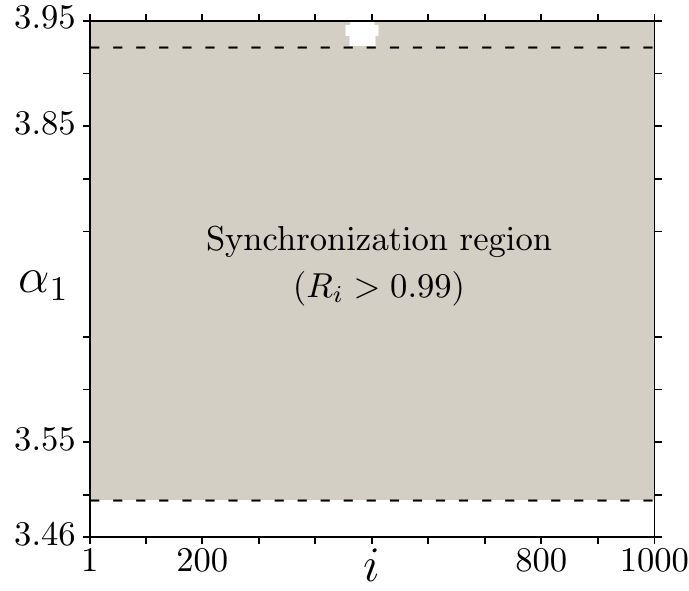}
    \caption{Region of external synchronization in the network (\ref{main_eq}) depending on the parameter $\alpha_1$ for $\gamma=0.4$, $\alpha_2=3.85$, $\sigma_1=0.23$, and $\sigma_2=0.15$ }
    \label{pic8_syncarea}
\end{figure}

We now turn to  a second case. We fix the control parameter in the driven network as $\alpha_2=3.85$ and vary $\alpha_1$ in the driver at $\gamma=0.4$. When $\alpha_1$ changes, the first ring  $x_i^t$  exhibits various spatiotemporal structures which are essentially different from the pattern observed in the second ring $y_i^t$ without inter-coupling. When the two rings are coupled, the response network shows  spatiotemporal structures which are synchronous with those ones realized in the driver network at given values of $\alpha_1$. 
In other words, the unidirectional inter-coupling causes the driven network to fully  follow and repeat the behavior of the driver network. 
Let us clarify this effect in more detail. Figure~\ref{pic8_syncarea} depicts the synchronization region for the fixed $\alpha_2=3.85$ and $\gamma=0.4$ and when $\alpha_1$ is varied. It is seen that the cross-correlation coefficient $R_i$ is larger than $0.99$ inside a finite range of $\alpha_1$ values and this clearly indicates the existence of external synchronization. The synchronization region is bounded by the lines $\alpha_1=3.5$ and $\alpha_1=3.9$. Outside this region (white color domain) the spatiotemporal structures are desynchronized and $R_i<1$.

%%%%%%%%Figure 8%%%%%%%%%%%%%%%%%%%%%%%%%%%%%%%%%%%%%%%%%
%%%%%%%%%%%%%%%%%%%%%%%%%%%%%%%%%%%%%%%%%%%%%%%

\begin{figure}[htbp]
    \centering
    
    \includegraphics[width=0.21\textwidth]{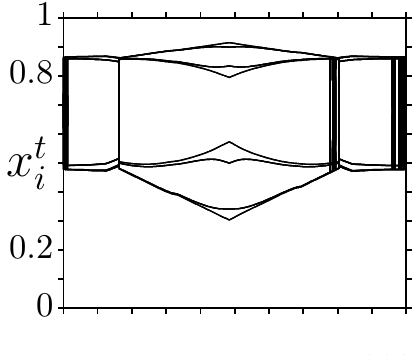}
    \includegraphics[width=0.21\textwidth]{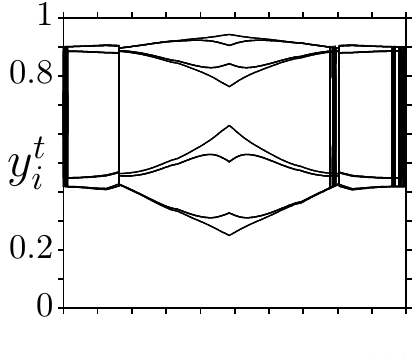}
    \\[-6.3em]~\hspace{25em}(a)
    \\[4.1em] \centering
    
    \includegraphics[width=0.21\textwidth]{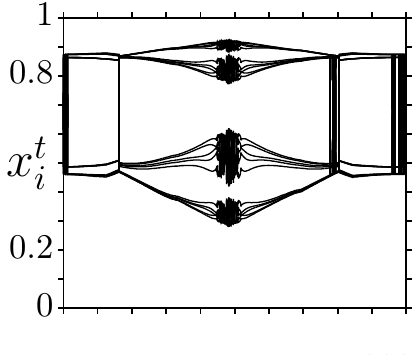}
    \includegraphics[width=0.21\textwidth]{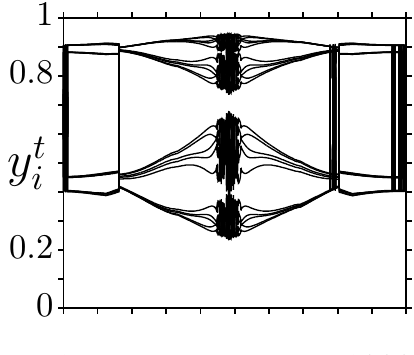}
    \\[-6.3em]~\hspace{25em}(b)
    \\[4.1em] \centering
    
    \includegraphics[width=0.21\textwidth]{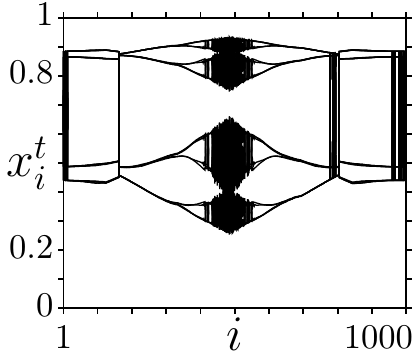}
    \includegraphics[width=0.21\textwidth]{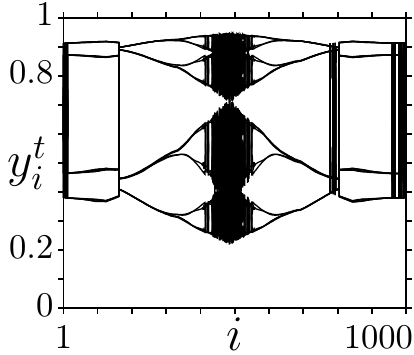}
    \\[-6.3em]~\hspace{25em}(c)
    \\[4.5em] \caption{Synchronous spatiotemporal structures in the coupled rings  $x_i^t$ (left) and $y_i^t$ (right) for different values of $\alpha_1$: $3.66$ (a), $3.7$ (b), $3.75$ (c) and for $\gamma=0.4$, $\alpha_2=3.85$, $\sigma_1=0.23$, and $\sigma_2=0.15$}
    \label{pic9_different_strs}
\end{figure}

As already mentioned, different values of  $\alpha_1$ in the driving network are associated with various spatiotemporal patterns which induce identical synchronous structures in the response network. Figure~\ref{pic9_different_strs} exemplifies typical structures which are realized  in both rings 
$x_i^t$ and $y_i^t$ for different values of $\alpha_1$. Indeed, when $\alpha_1$ is varied, the spatiotemporal patterns change their form but  remain identical and synchronous inside the synchronization region.

%%%%%%%%Figure 9%%%%%%%%%%%%%%%%%%%%%%%%%%%%%%%%%%%%%%%%%
%%%%%%%%%%%%%%%%%%%%%%%%%%%%%%%%%%%%%%%%%%%%%%%%%

\begin{figure}[htbp]
    \centering
    \includegraphics[width=0.37\textwidth]{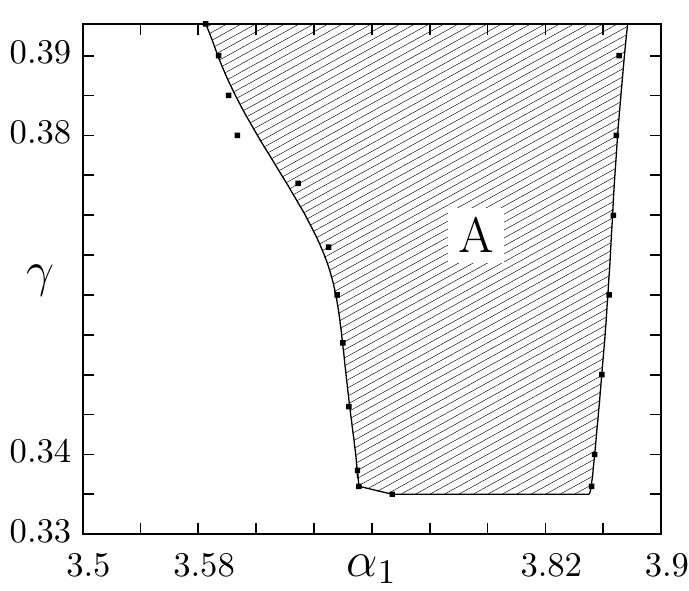}
    \caption{Region of external synchronization of spatiotemporal structures in the network (\ref{main_eq}) in the ($\alpha_1,\gamma$) parameter plane for $\alpha_2=3.85$, $\sigma_1=0.23$, and $\sigma_2=0.15$}
    \label{pic10_syncarea}
\end{figure}

Our numerical studies have shown that when varying $\alpha_1$ the width of synchronization region depends on the inter-coupling strength $\gamma$. To illustrate this dependence, the region of external synchronization  of spatiotemporal structures in the model (\ref{main_eq}) is depicted in Fig.~\ref{pic10_syncarea} in the $(\alpha_1,\gamma)$ parameter plane.

%%%%%%%%Figure 10%%%%%%%%%%%%%%%%%%%%%%%%%%%%%%%%%%%%%%%%%
%%%%%%%%%%%%%%%%%%%%%%%%%%%%%%%%%%%%%%%%%%%%%%%%%

As can be clearly seen from the figure, the synchronization region has a tongue-like form, is  characterized by a certain threshold value $\gamma_{\rm{th}}\approx0.333$ and expands in the parameter $\alpha_1$ as the inter-coupling $\gamma$ increases.

\section{Mutual synchronization}

We  return to the network (\ref{main_eq}) with the initially chosen parameter values: $\alpha_1=3.7$, $\alpha_2=3.85$, $P=R=320$, $N=1000$ and set the equal value of nonlocal coupling strengths in both rings as $\sigma_1=\sigma_2=0.28$.  We now intend to explore the mutual synchronization between the two coupled rings. For this purpose we introduce the symmetric (bidirectional) inter-coupling between the ensembles $\gamma_{12}=\gamma_{21}=\gamma$ and analyze numerically the evolution of spatiotemporal structures in the network  (\ref{main_eq})  when the inter-coupling strength is  varied in the interval $[0, 0.8]$. We iterate the network (\ref{main_eq}) during $2\times 10^5$ and discard the first $10^6$ iterations as a transient.

%  При указанных значениях параметров в ансамблях $x_i^t$ и $y_i^t$ в отсутствии связи $\gamma_{12}=\gamma_{21}=0$ будут реализовываться различные пространственно-временные структуры, так как введена расстройка по параметру $\alpha$ ($\alpha_1\not=\alpha_2$). 
%Введем симметричную связь $\gamma_{12}=\gamma_{21}=\gamma$ и проведем расчеты реализуемых структур в интервале $0\leqslant\gamma\leqslant0.8$.

\begin{figure}[htbp]
    \centering
    \includegraphics[width=0.18\textwidth]{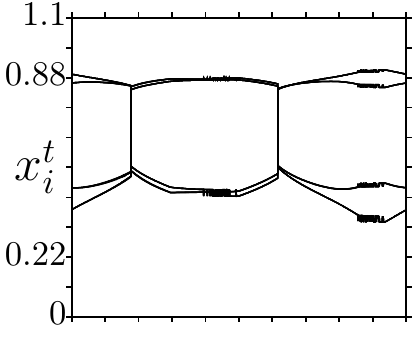}~\hspace{-2.0em}
    \includegraphics[width=0.18\textwidth]{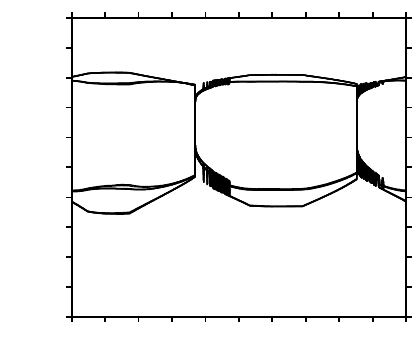}~\hspace{-2.0em}
    \includegraphics[width=0.18\textwidth]{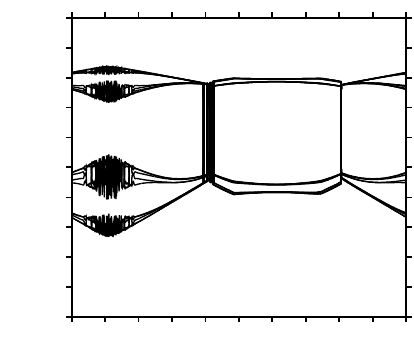}
    \\[-6pt]
    \includegraphics[width=0.18\textwidth]{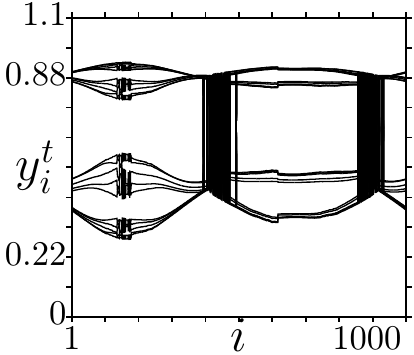}~\hspace{-2.0em}
    \includegraphics[width=0.18\textwidth]{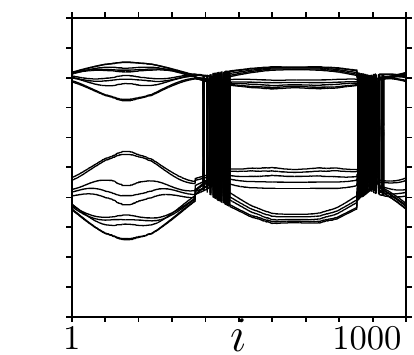}~\hspace{-2.0em}
    \includegraphics[width=0.18\textwidth]{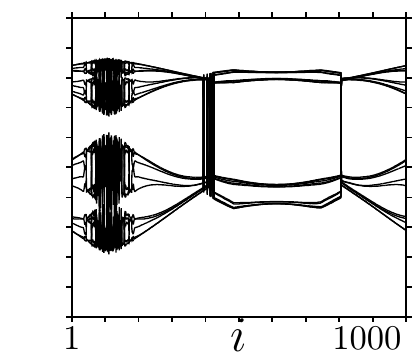}
    \\[-2pt]~\hspace{1.2em}(a)\hspace{7.0em}(b)\hspace{7.0em}(c)
    \\[-.3em] \caption{Mutual synchronization of spatiotemporal structures in symmetrically coupled ensembles   $x_i^t$ (upper panel) and $y_i^t$ (low panel) for different values of inter-coupling strength $\gamma$: $0.011$ (a), $0.025$ (b), and $0.075$ (c) at $\alpha_1=3.7$, $\alpha_2=3.85$, $\sigma_1=\sigma_2=0.28$}
    \label{pic11_gotosync}
\end{figure}

\begin{figure}[htbp]
    \centering
    \includegraphics[width=0.37\textwidth]{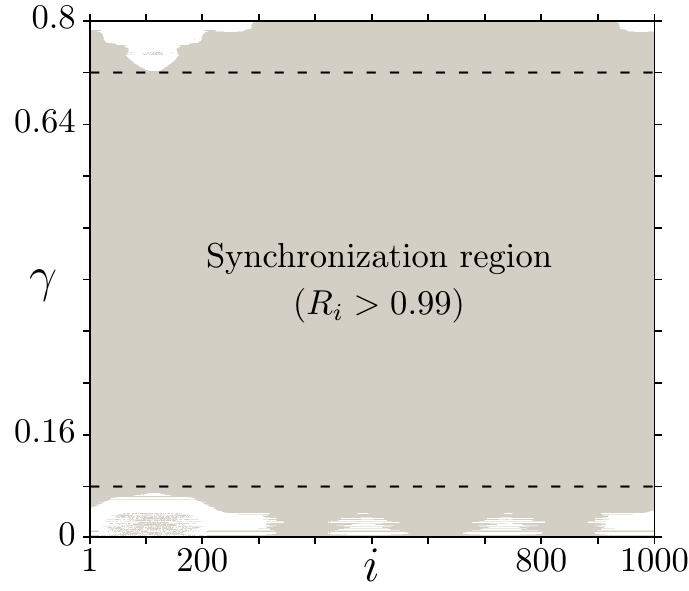}
    \caption{Region of mutual synchronization of spatiotemporal strcutures in the network ~(\ref{main_eq}) when varying the inter-coupling $\gamma$ and for $\alpha_1=3.7$, $\alpha_2=3.85$, and $\sigma_1=\sigma_2=0.28$}
    \label{pic12_syncarea_gamma}
\end{figure}

The calculation results are presented in Fig.~\ref{pic11_gotosync} where spatiotemporal profiles for the coordinates of each sub-network are shown for three different values of the inter-coupling strength. As can be seen from Fig.~\ref{pic11_gotosync},c, the spatiotemporal patterns are synchronized at $\gamma=0.075$ and this is corroborated by calculating the cross-correlation coefficient $R_i$ which is larger than $0.99$ in this case.  Thus,
the mutual synchronization is established in the network (\ref{main_eq}).
It is important to point out that the structures $x_i^t$ and $y_i^t$ synchronized at $\gamma=0.075$ differ from the  corresponding patterns which are realized in these ensembles when there is no coupling between them.  The effect of mutual synchronization takes place within a finite region of synchronization, which is depicted in Fig.~\ref{pic12_syncarea_gamma}. 
As follows from this figure, the cross-correlation coefficient $R_i\geq 0.99$ inside a certain bounded range of the inter-coupling strength,   $0.075\leq\gamma\leq 0.72$, and thus indicates the existence of a finite region of mutual synchronization of spatiotemporal structures in the network (\ref{main_eq}). 

%%%%%%%Figure 11%%%%%%%%%%%%%%%%%%%%%%%%%%%%%%%%%%%%%%%%%
%%%%%%%%%%%%%%%%%%%%%%%%%%%%%%%%%%%%%%%%%%%%%%%

%%%%%%%%Figure 12%%%%%%%%%%%%%%%%%%%%%%%%%%%%%%%%%%%%%%%%%
%%%%%%%%%%%%%%%%%%%%%%%%%%%%%%%%%%%%%%%%%%%%%%%%%

\section{Discussion and conclusions}
\label{conclusions}

In our work we have studied numerically the dynamics and synchronization phenomena in the two layer network of nonlocally coupled chaotic discrete-time systems (\ref{main_eq}) with unidirectional and symmetrical couplings between the one layer networks. Our results have confidently indicated that the effects of external and mutual synchronization of spatiotemporal structures, including chimera states, take place in the considered network. We have demonstrated that synchronized patterns are identical and  observed within finite regions of synchronization, which are the neccesary and sufficient conditions for realizing and justifying the synchronization effect. 
The identity of synchronous structures and the synchronization regions  are estimated quantitatively by calculating the cross-correlation coefficient  $R_i$~(\ref{correlations_eq}) of oscillations of the $i$th elements of the interacting ensembles, where  $i=1,2,\dots,N$.
 
 The  phenomenon of synchronization of spatiotemporal structures can be qualitatively compared with the classical effect of synchronization of periodic self-sustained oscillations \cite{PIK01,Springer07}. In this case the spectral oscillation line at  a frequency $\omega$ can be treated as the simplest structure. In the case of external synchronization the behavior of a driver oscillator is characterized by the spectral line at the frequency $\omega_1$ and the structure of a driven oscillator -- by the spectral line at the frequency $\omega_0$. When the external synchronization is established, a frequency locking occurs inside the synchronization region. This means that the driven oscillator frequency $\omega_0$ shifts and coincides with the driver oscillator frequency $\omega_0=\omega_1$, e.g., $\omega_0$ is locked by $\omega_1$. This frequency equality remains unchanged (constant) within the synchronization region. Varying  $\omega_1$ causes  $\omega_0$ to change so that the equality $\omega_0=\omega_1$  always holds inside the synchronization region. A similar effect takes place for mutual synchronization between two oscillators with different natural frequencies $\omega_1$ and $\omega_2$. The only  difference consists in the fact that inside the synchronization region, the oscillations are generated either with frequency $\omega_1$ or $\omega_2$, or with a certain intermediate frequency   $\omega_1<\omega<\omega_2$. 

Our numerical results presented in the paper show that a qualitatively equivalent behavior is realized for the two layer network of nonlocally coupled nonlinear discrete-time systems. In the case of external synchronization the spatiotemporal structure of the response network "is locked" by the structure of the driver network and the identical patterns remain unchanged within the synchronization region (see Figs.~ \ref{pic4_synced}--\ref{pic7_syncarea}). When the mutual synchronization takes place, the spatiotemporal structures in the coupled one layer networks are mutually locked. Moreover, the resulting synchronous structures differ from the initially observed patterns which  are realized in the coupled rings without inter-coupling   (see Fig.~\ref{pic11_gotosync}). This peculiarity is also very typical for mutual synchronization of coupled periodic self-sustained oscillators.

Finally,  the region of external synchronization pictured in Fig.~\ref{pic10_syncarea} is the most illustrative and bright result of our studies. This plot can be qualitatively compared with  relevant pictures  obtained in the framework of the classical theory of external synchronization of periodic self-sustained oscillations \cite{PIK01,Springer07}. As it is well known, the region of external synchronization is constructed on the "external signal amplitude -- frequency mismatch (detuning)" parameter plane. The synchronization region narrows with decreasing external amplitude and vanishes at its zero value.  In our case the inter-coupling strength $\gamma$ plays the role of the external signal amplitude and defines the oscillation amplitude (the coordinate value) $y_i$ in the response network. At a fixed $\alpha_2=const$, the parameter $\alpha_1$ determines a detuning (mismatch) of structures, $\Delta\alpha=|\alpha_1-\alpha_2|$. However, unlike the external synchronization of periodic self-sustained oscillations, there is a synchronization threshold in the inter-coupling strength $\gamma$ when the two one layer networks are unidirectionally coupled (see Fig.~\ref{pic10_syncarea}). We believe that this threshold effect is associated with a complicated topology of the inner couplings between the network  (\ref{main_eq}) elements. 

The above comparison enables one to  conclude that the results of this paper can be considered as a generalization of the notions of the classical phenomenon of  synchronization of periodic self-sustained oscillations to the case of synchronization of spatiotemporal structures in multilayer networks of nonlocally coupled nonlinear oscillators. We also beleive that our findings will considerably contribute to the synchronization of chimera states and various spatiotemporal structures in multilayer networks.

\section{Acknowledgments}
This work was supported by DFG in
the framework of SFB 910 and by the Russian Ministry of Education and Science (Project Code 3.8616.2017/8.9).

%\bibliographystyle{apsrev}
%\bibliography{bibliography}

\end{document}